\documentclass{aa}         % (traditional abstract) 

\usepackage{graphicx}
\usepackage{natbib}
\bibpunct{(}{)}{;}{a}{}{,} % to follow the A&A style

\usepackage{verbatim}
\usepackage{txfonts}
\usepackage[figuresright]{rotating}
\usepackage{url}

\begin{document}
   \title{The Pisa Stellar Evolution Data Base for low-mass stars}

%.G.   \subtitle{}

   \author{M. Dell'Omodarme \inst{1}, G. Valle \inst{1}, 
S. Degl'Innocenti \inst{1,2}, P.G. Prada Moroni \inst{1,2}
          }

   \authorrunning{Dell'Omodarme,M. et al.}

   \institute{Dipartimento di Fisica ``Enrico Fermi'',
Universit\`a di Pisa, largo Pontecorvo 3, Pisa I-56127 Italy
\and
  INFN,
 Sezione di Pisa, Largo B. Pontecorvo 3, I-56127, Italy}

   \offprints{G. Valle, valle@df.unipi.it}

   \date{Received 12/12/2011; accepted 26/01/2012}

  \abstract
   {   The last decade showed an impressive observational effort from the
       photometric and spectroscopic point of view for ancient stellar
       clusters in our Galaxy and beyond, leading to important and sometimes
       surprising results.
}
   {   The theoretical interpretation of these new observational results
       requires updated evolutionary models and isochrones spanning a wide
       range of chemical composition so that the possibility
       of multipopulations inside a stellar cluster is also taken also into account.
}
{   With this aim we built the new ``Pisa Stellar Evolution Database'' of
  stellar 
    models and isochrones by adopting a well-tested evolutionary
    code (FRANEC) implemented with updated physical and chemical inputs. 
    In particular, our code adopts realistic atmosphere models and an updated 
    equation of state, nuclear reaction rates and opacities calculated with
    recent solar elements mixture.
}
  {   A total of 32646 models have been computed in the range of initial
      masses $0.30\div1.10$ $M_{\sun}$ for a grid of 216 chemical compositions
      with the fractional metal abundance in mass, $Z$, ranging from 0.0001
      to 0.01, and the original helium content,
      $Y$, from 0.25 to 0.42.  Models were computed for both solar-scaled and
       $\alpha$-enhanced abundances
      with different external convection efficiencies. Correspondingly,
      9720 isochrones were computed in the age range $8\div15$ Gyr, in
      time steps of 0.5 Gyr. The whole database is available to the
      scientific community on the web.
Models and isochrones were compared with recent
      calculations available in the literature and with the color-magnitude
      diagram of selected Galactic globular clusters.
 The dependence of relevant
      evolutionary  
      quantities, namely turn-off and horizontal branch
      luminosities, on the 
      chemical composition and 
      convection efficiency were analyzed in a quantitative statistical
      way and analytical formulations were made available for reader's
      convenience.  
These relations can be useful in several fields of stellar
evolution, e.g. evolutionary properties of binary systems, synthetic
models for simple stellar populations and for star counts in galaxies,
and chemical evolution models of galaxies. 
 }
    { 
}


   \maketitle

\section{Introduction}\label{sec:intro}

Globular clusters (GCs) are of fundamental relevance for our knowledge of the
Universe. They are among the most ancient objects in galaxies and consequently can
help to understand galaxies evolution and constrain the age of the Universe,
moreover they are intrinsically bright objects that can be observed at far
distances.

Thanks to an impressive improvement of spectroscopic and photometric
observational capabilities, the last decade was a very exciting period for
globular cluster researches. Globular clusters cannot anymore be considered as ``simple
stellar populations'', i.e. as 
an assembly of coeval, chemically homogeneous stars. Recent
spectroscopical investigations 
\citep[see e.g.][and references therein]{carretta2010,  bragaglia2010, 
melendez2009, yong2008, smith2005, gratton2004} showed that every GC studied so
far hosts at least two different stellar generations, distinct in the
abundance of several elements (C, N, O, Na, Mg, etc..). The situation is
made additionally complex and interesting by an increasing number of discoveries
within the most massive globular clusters of multiple stellar populations,
photometrically distinct in the color-magnitude (CM) diagram
 \citep[see e.g.][]{pancino2011, pancino2000, 
carretta2009, villanova2007, norris2004}. 
Moreover, some of these
populations seem to show a very high original helium abundance (up to
$Y \approx$ 0.40) that is not accompanied by a corresponding increase in the
iron 
 abundance \citep[see e.g.][]{dupree2011, marino2011, marino2009, bellini2010,
milone2010, milone2008, anderson2009,  piotto2007, piotto2005}.  
The multiple-population
phenomenon in star clusters is not restricted to our Galaxy: high-precision
photometry observations show the presence of distinct populations inside old
clusters of the Magellanic Clouds 
\citep[see e.g.][]{milone2009, glatt2008, mackey2008, 
mackey2007}.

The theoretical interpretation of these data to recover the
evolutionary history of clusters requires updated tracks and isochrones
databases. They must span 
a wide range of chemical compositions with the inclusion of very high helium
abundances, to properly model the presence of multipopulations in old
clusters in the Milky Way and in near dwarf galaxies.

To this aim we developed a large, homogeneous database with a fine grid of
tracks and isochrones with 216 different chemical compositions, both solar-
scaled and $\alpha$-enhanced, calculated with the recent \citet{AGSS09}
solar elements mixture for different external convection efficiencies. 

Similar databases are present in literature: BaSTI \citep{teramo04,teramo06}, 
Dartmouth \citep{dotter07,dotter08}, and Padova STEV \citep{padova08,padova09}.
They differ among each other for the adopted chemical
compositions and physical inputs (opacities, atmospheric models, equations of
state, nuclear reactions rates, convection efficiencies etc.). 
Our models, computed with the current physical inputs, can therefore be compared
with other results to 
estimate the effects of the variation on chemical
composition and physical inputs. A comparison for the most relevant evolutionary
features is presented in this paper. 

All calculations are available to the astrophysical
community\footnote{\url{http://astro.df.unipi.it/stellar-models/}}.
At the same link an extended database for pre-main
sequence (PMS) stars with different chemical compositions is already available, as described in a
previous paper \citep{Tognelli2011}.

As a check, our models are compared with the CM diagram of
three selected Galactic globular clusters spanning the metallicity range
of GCs in the Milky Way and in the Magellanic Clouds (from [Fe/H] = -2.35 to
[Fe/H]  = -0.76).  

Section \ref{sec:franec} is devoted to a short description of the
physical inputs adopted in our evolutionary code, Sect.
\ref{sec:comparison} presents the comparison with the selected globular
clusters, Sect. \ref{sec:db-descr} is devoted to the description of our
database and Sect. \ref{sec:db-comparison} shows the comparison with other
selected stellar evolution model databases available 
in the literature. In Section \ref{sec:modelfit} the dependence of relevant
evolutionary quantities, namely the turn-off (TO) and the horizontal branch
(HB) luminosities, on the chemical composition and convection efficiency were
analyzed in  
a quantitative statistical way and analytical formulations were made
available for reader's convenience. 
The concluding remarks
are given in Section \ref{sec:concl}.

\section{Input physics for evolutionary models}\label{sec:franec}

The adopted stellar evolutionary code, FRANEC, has been extensively described 
in previous
papers \citep[and references therein]{cariulo04,scilla2008}, while recent
updates of the physical inputs are discussed in \citet{cefeidi} and
\citet{Tognelli2011}. We include  
here only a brief description of the adopted physical inputs, pointing out the
updates relevant for low-mass model evolution. Present physical and chemical
inputs are summarized in Table \ref{table:2}, where a comparison with other 
available databases is also reported (see also Sec.~\ref{sec:db-comparison}).

Present calculations used the most recent version of the OPAL equation
of state, EOS,\footnote{Tables available at 
  \url{http://www-phys.llnl.gov/Research/OPAL/}} 2006
\citep{rogers96, rogers02}. 

For temperatures higher than $10^4$ K
radiative opacities were taken from the OPAL group
\citep{rogers96}\footnote{\url{  
http://opalopacity.llnl.gov/}} in the version released in 2006, so
that high-temperature opacities and EOS are fully consistent,
whereas for lower temperatures the code adopts molecular opacities by
\citet{ferg05}\footnote{\url{http://webs.wichita.edu/physics/opacity/}}.
In both cases opacity tables are computed for the solar mixture by 
\citet{AGSS09},
both solar-scaled and $\alpha$-enhanced with [$\alpha$/Fe] = 0.3.
For electron conduction opacities we adopted the recent results by 
\citet{casspote07}, based on \citet{pote99a}.

Nuclear reaction rates were taken from the NACRE compilation \citep{nacre}
except for 
$^{12}$C$(\alpha,\gamma)^{16}$O and 
$^{14}$N$(p,\gamma)^{15}$O, for which we adopted more 
recent estimates, by \citet{12c} and \citet{14n} respectively;
the $^{3}$He$(\alpha,\gamma)^{7}$Be reaction rate was taken from
\citet{cyburt08}. 
The energy losses by plasma neutrinos were taken from \citet{haft94}, while
for the  
other neutrino emission processes we refer to \citet{itoh96}.

For convective mixing, we adopted the Schwarzschild criterion to define regions
in which convection elements are accelerated. Semiconvection during the
central He-burning phase 
\citep{castellani1971} was treated following the numerical scheme
described in \citet{castellani1985}. Breathing pulses were suppressed
\citep{cassisi2001, castellani1985} following the procedure
suggested by \citet{caputo1989}. 

To model external convection we adopted, as usual, the mixing length
formalism \citep{bohmvitense58} in which the convection
efficiency is  parametrized in terms of the mixing length parameter
$\alpha_{ml}$ 
i.e. the ratio between the mixing length and the local pressure scale height:
$\alpha_{ml}=l/H_p$. 

Present models include realistic atmospheric models by \citet{brott05} 
(hereafter BH05), computed using the PHOENIX code
\citep{hauschildt99,hauschildt03},  
available in the range $3000 \; {\rm K} \le T_{eff} \le 10000 \;
{\rm K}$, $0.0 \le \log g \; {\rm (cm \; s^{-2})} \le 5.0$, and $-4.0 \le {\rm
  [M/H]} 
\le 0.5$. The mixing length scheme was adopted to describe the convection with
$\alpha_{ml}$ = 2.0. 
In the range $10000 \; {\rm K} \le T_{eff} \le 50000 \; {\rm K}$, $0.0
\le \log g \; {\rm (cm \; s^{-2})} \le 5.0$, and $-2.5 \le {\rm [M/H]} 
\le 0.5$, where models from BH05 are unavailable, we used   
models by \citet{castelli03} (hereafter CK03). In this case the
mixing lenght adopted is $\alpha_{ml}$ = 1.25. A discussion of the
influence of the different mixing values and of the solar mixture adopted in
the atmospheric model can be found in \citet{Tognelli2011}.

Atomic diffusion was included, taking into account the
effects of gravitational settling and thermal diffusion with
 coefficients given by \citet{thoul94}. Radiation-driven diffusion
acceleration \citep[see e.g.][]{richer98,richard02} and rotation \citep[see
  e.g.][]{palacios2003, maeder1998}  
are not included in the models.

\section{Comparison with observational data }\label{sec:comparison}

As a check of our models, we compared them with three well known, not too
heavily reddened, globular clusters that span a wide range of metallicity
values: M92, M3 and 47 Tuc.  We selected
M92 as an example of the most metal-poor clusters ([Fe/H] = -2.35, see
\citealt{carretta2009}, $Z=0.0001$) taking the photometric data from
\citet{dicecco2010}, M3 as moderately metal-rich cluster ([Fe/H] = -1.50, see
\citealt{carretta2009}, $Z=0.0007$), data taken from \citet{rey2001}, and 47 Tuc
as 
metal-rich cluster ([Fe/H] = -0.76 see \citealt{carretta2009}, $Z=0.004$), data
taken from \citet{Bergbusch2009}.
For M92 good quality data are available both
for ($V$, $B-V$) and ($V$, $V-I$) diagrams.  
The quoted cluster metallicities were
obtained from the observed [Fe/H] values by adopting as a reference the heavy
elements 
 solar mixture by \citet{AGSS09} and an enhancement of the $\alpha$-elements
  [$\alpha$/Fe] = 0.3. The required initial helium
abundance was obtained by assuming the recent value of the primordial helium
abundance  
$Y_p= 0.2485$ and the helium-to-metal enrichment ratio, $\Delta
Y/\Delta Z = 2$, as described in greater detail in Sec.~\ref{sec:db-descr}.

Following a widely adopted procedure, the mixing length parameter
$\alpha_{ml}$ was  
calibrated by reproducing the red giant branch (RGB) color. This result is
also dependent on the atmospheric models adopted to transform evolutionary
calculations from the theoretical ($\log L - \log T_{eff}$) to the
observational plane. We adopted the synthetic spectra provided by 
\citet{brott05} for $T_{eff} \le 10000$ K 
and by \citet{castelli03} for $T_{eff} > 10000$ K.

\begin{figure*}
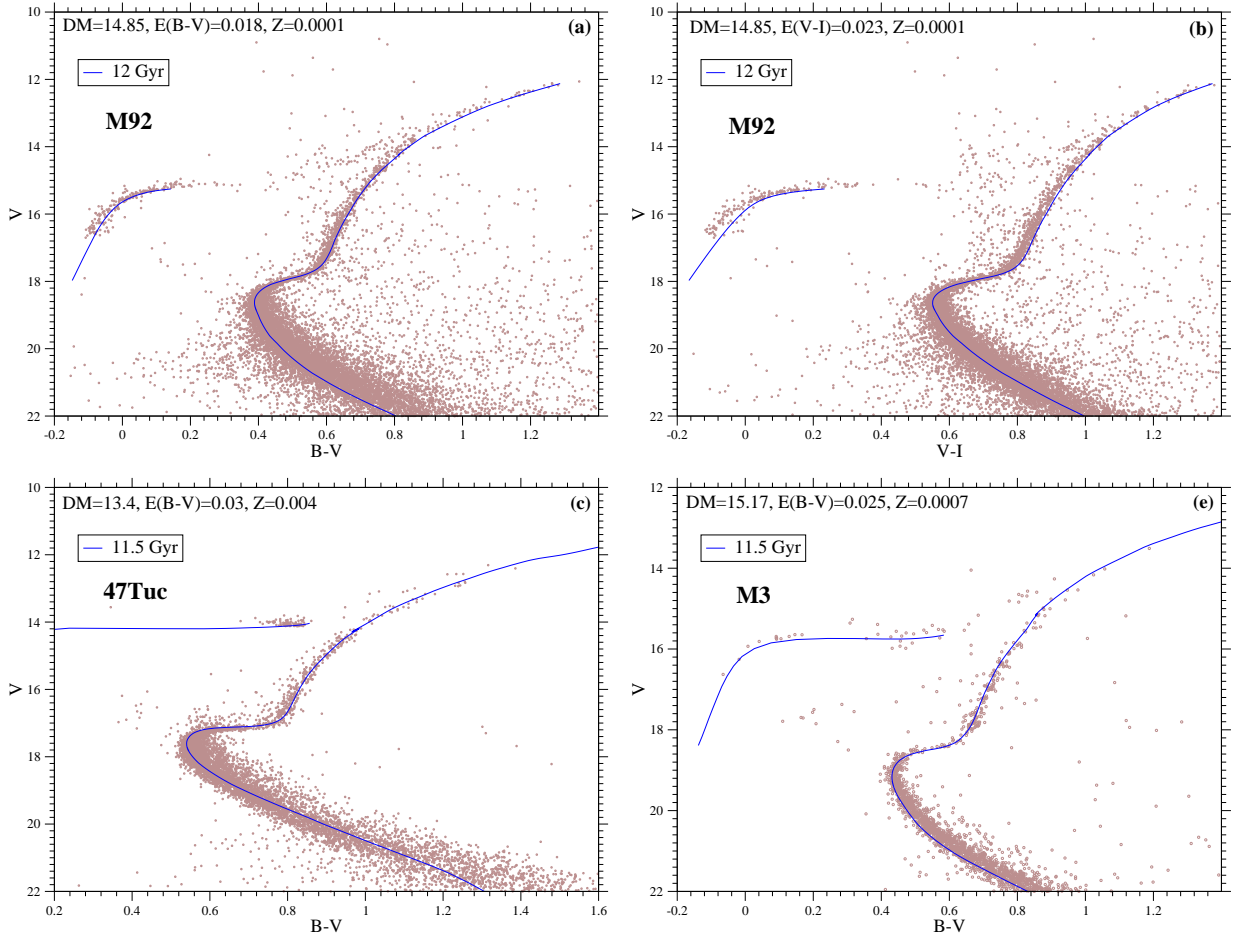

\centering
\includegraphics[width=8cm]{M92_VB-V_DM.eps}\hspace{2mm} 
\includegraphics[width=8cm]{M92_VI-V_DM.eps}\\\vspace{2mm}
\includegraphics[width=8cm]{47Tuc_VB-V.eps}\hspace{2mm}
\includegraphics[width=8cm]{M3_VB-V.eps}
\caption{Comparison of present isochrones and CM diagrams for three
selected globular clusters, see text. For each cluster the estimated values
for age, distance modulus and reddening are labeled together with the
metallicity, as calculated from the observed [Fe/H] value 
(see Sec.~\ref{sec:comparison}). Top left and top right panel:  
comparison for the globular cluster M92, in the $(V,B-V)$ and in the ($V$,
$V-I$) 
diagram, respectively. Bottom right panel: the same, but for the M3 cluster in the
($V$, $B-V$) diagram. Bottom
left panel: the same, but for 47 Tuc cluster in the
($V$, $B-V$) diagram.
}
\label{fig:M92andM3and47Tuc}
\end{figure*}

Figure \ref{fig:M92andM3and47Tuc} shows the very good agreement between
theory and observations for the selected clusters in the ($V$, $B-V$) and
($V$, $V-I$) filters. In all examined cases, the best concordance
is achieved
for $\alpha_{ml}$ = 1.90. The inferred values for the cluster
parameters (age, distance modulus, and reddening) are reported in the
figure. Even if our
purpose is only to check the general agreement between the present set of
models and data, we note that our
estimates for age, distance modulus and reddening are consistent, within
the uncertainties, with the recent ones available in the literature 
(see e.g. \citealt{dicecco2010, kraft2003, salaris2002, vandenBerg2002} 
for M92; \citealt{kraft2003, rey2001, Yi2001} for M3; 
\citealt{Bergbusch2009,
percival2002, grundahl2002} and references therein, \citealt{ 
zoccali2001} for 47 Tuc). 
We are aware that very high quality
photometric data for 47 Tuc show the possible presence of multipopulation
from the analysis of the subgiant branch \citep[see e.g.][]{anderson2009},
however, a discussion of this problem is beyond the scope of the
present paper.

\section{Database description: stellar tracks and Isochrones}
\label{sec:db-descr}

Stellar tracks were computed from the PMS phase through the evolution
of the whole H and He burning phases up to the first thermal
pulse, except for the lowest masses, which take longer than the Hubble time to
exhaust 
the central hydrogen.  We covered a range
of masses from 0.30 $M_{\sun}$ to 1.10 $M_{\sun}$, in steps of 0.05
$M_{\sun}$.  The limit of 0.30
$M_{\sun}$ was chosen because lower masses present along its evolution
temperature and pressure value not covered by the OPAL EOS.  As
shown in Table \ref{table:1db}, we selected 19 metallicity values, with $Z$
varying from $Z=0.0001$ to $Z=0.01$.
 
For each $Z$ value, we computed models with six different helium
abundances. Five of 
them are fixed values ($Y$ = 0.25, 0.27, 0.33, 0.38, 0.42) that simulate
different helium enrichments up to the very high values supposed for
some stellar samples in multipopulation clusters \citep{lee05,
villanova2007, piotto2007, piotto09}, while the last one follows the often
adopted 
linear helium-to-metal enrichment law given by:
$Y=Y_p+\frac{\Delta Y}{\Delta  Z}Z$.
For the cosmological $^4$He abundance we adopted the value $Y_p=0.2485$,
as recently estimated by WMAP 
\citep{cyburt04,steigman06,peimbert07a,peimbert07b}.
For the galactic helium-to-metal enrichment ratio we chose 
$\Delta Y$/$\Delta Z=2$, 
a typically assumed value for this quantity that is still affected by several important
sources of  
uncertainty \citep{pagel98,jimenez03,flynn04,gennaro10}.

We adopted the solar heavy-element mixture recently provided by \citet{AGSS09}. 
We computed models also for an enhanced abundance of the $\alpha$ elements
 with respect to the solar mixture with [$\alpha$/Fe] = 0.3 
\citep[see e.g. the discussion in][]{ferraro1999}. 
\citet{salaris1993} showed that $\alpha$-enhanced
models can be reproduced by the solar-scaled ones with the same total
metallicity, provided that the ratio of the high (C, N, O, Ne) over the low 
(Mg, Si, S, Ca, Fe) ionization potential elements is preserved. However, as
pointed out by several authors \citep[see e.g.][]{vandenBerg2000, salaris1998,
weiss1995},   
this property starts to break
down at Z $\approx$ 0.002 and becomes less and less reliable with increasing metallicity. 

As already illustrated (see Section
\ref{sec:comparison}), the value $\alpha_{ml}=1.90$ was
calibrated against the observed CM diagrams
of the globular clusters M92, M3 and 47Tuc. However, since the effective
temperatures of low-mass stars are considerably affected by changes in the
$\alpha_{ml}$ value, we performed calculations also for two other values of the
mixing length parameter, $\alpha_{ml}$ = 1.70; 1.80. 
The solar-calibrated mixing length parameter 
  is $\alpha_{ml}$ = 1.74.
The
mixing length
is merely a fitting parameter, linked to the still unavoidable uncertainties
in external convection efficiency calculations. For these reasons the
required $\alpha_{ml}$ value could be, in principle, different not only for
different stellar masses and chemical compositions, but also for different
evolutionary phases of the same model \citep[see e.g.][]{brocato1999}. 
Fortunately its influence
on model luminosities is quite negligible \citep[see e.g.][]{chaboyer1995} 
for reasonable values of this quantity.

For each set of parameters,
two types of track files were included.
The first group contains the output of the calculations beginning from the PMS
and ending 
either at the helium flash (for $M \geq 0.55 M_{\sun}$)  
or at central hydrogen exhaustion ($M \leq 0.50 M_{\sun}$).
The second group, computed for each calculation reaching the helium flash in
less than 15 Gyr, consists of files 
beginning from the zero-age horizontal branch (ZAHB) model and 
ending at the onset of thermal pulses. 
Table \ref{table:1db} summarizes the
models included in the database.  Isochrones were computed in the
typical GC age range, from 8 to 15 Gyr, with time steps of
0.5 Gyr. This part of the database therefore contains a total of 11016 tracks
starting from PMS, 
5549 tracks starting from ZAHB, and 9720 isochrones.

\begin{table*}
\caption{Summary of calculated data base tracks (see Sec.\ref{sec:db-descr})}
\label{table:1db}      
\centering          
\begin{tabular}{c|c|c|c|c|c|c}     
\hline\hline  
\multicolumn{7}{l}{Mass range: from 0.30 $M_{\sun}$ to 1.10 $M_{\sun}$, steps
  of 0.05 $M_{\sun}$} \\ 
\multicolumn{7}{l}{Mixture: \citet{AGSS09}, [$\alpha$/Fe] = 0.0, 0.3}\\
\multicolumn{7}{l}{$\alpha_{ml}$ = 1.70, 1.80, 1.90}\\
\hline
$Z$ & \multicolumn{6}{c}{$Y$}\\
\hline 
   0.0001 & 0.249 & 0.250 & 0.270 & 0.330 & 0.380 & 0.420\\  
   0.0002 & 0.249 & 0.250 & 0.270 & 0.330 & 0.380 & 0.420 \\
   0.0003 & 0.249 & 0.250 & 0.270 & 0.330 & 0.380 & 0.420 \\
   0.0004 & 0.249 & 0.250 & 0.270 & 0.330 & 0.380 & 0.420   \\
   0.0005 & 0.250 & 0.250 & 0.270 & 0.330 & 0.380 & 0.420   \\
   0.0006 & 0.250 & 0.250 & 0.270 & 0.330 & 0.380 & 0.420\\
   0.0007 & 0.250 & 0.250 & 0.270 & 0.330 & 0.380 & 0.420\\
   0.0008 & 0.250 & 0.250 & 0.270 & 0.330 & 0.380 & 0.420\\
   0.0009 & 0.250 & 0.250 & 0.270 & 0.330 & 0.380 & 0.420\\
   0.0010 & 0.250 & 0.250 & 0.270 & 0.330 & 0.380 & 0.420\\
   0.0020 & 0.252 & 0.250 & 0.270 & 0.330 & 0.380 & 0.420\\
   0.0030 & 0.254 & 0.250 & 0.270 & 0.330 & 0.380 & 0.420\\
   0.0040 & 0.256 & 0.250 & 0.270 & 0.330 & 0.380 & 0.420\\
   0.0050 & 0.258 & 0.250 & 0.270 & 0.330 & 0.380 & 0.420\\
   0.0060 & 0.260 & 0.250 & 0.270 & 0.330 & 0.380 & 0.420\\
   0.0070 & 0.262 & 0.250 & 0.270 & 0.330 & 0.380 & 0.420\\
   0.0080 & 0.264 & 0.250 & 0.270 & 0.330 & 0.380 & 0.420 \\
   0.0090 & 0.266 & 0.250 & 0.270 & 0.330 & 0.380 & 0.420\\
   0.0100 & 0.268 & 0.250 & 0.270 & 0.330 & 0.380 & 0.420\\
\hline  
\end{tabular}
\end{table*}

Track files names were chosen to clearly indicate
the inputs used in the simulation. As an example, for $M$ = 0.80 $M_{\sun}$,
$Z$ = 0.01, $Y$ = 0.25, $\alpha_{ml}$ = 1.90, solar-scaled Asplund 2009 mixture,
the track file from PMS to RGB flash is named {\it
  OUT\_M0.80\_Z0.01000\_He0.2500\_ML1.90\_AS09a0.DAT}. Outputs starting from
ZAHB are named appending the value of ZAHB mass to the name:
{\it
  OUT\_M0.80\_Z0.01000\_He0.2500\_ML1.90\_AS09a0\_ZAHB0.8000.DAT}. 
In each file, the
following quantities are listed: model number, 
age in $\log$ age (yr), 
luminosity in $\log L/L_{\sun}$, 
effective temperature in $\log T_{eff}$,
central temperature in $\log T_{c}$, 
central density in $\log \rho_{c}$, 
mass of the helium core ($M_{He}^c/M_{\sun}$),
mass of the star ($M_{\sun}$), 
fractional central abundance in mass of hydrogen 
(after the H exhaustion: fractional central abundance in mass of He),
luminosity of the $pp$ and of the CNO chains, 
luminosity of the $3 \alpha$ burning,
luminosity of the gravitational energy,
radius of the star ($R_{\sun}$),
logarithm of surface gravity. 
Although in our models the mass is constant, the mass of the evolving
  star is included to allow a possible 
  future inclusion of mass loss in the database without changing the layout
  of the output tables. 
Additional evolutionary quantities for the calculated models are
available on request.

Besides the models presented above, a
grid of HB models were calculated from an unique RGB
progenitor mass for each chemical composition (without mass loss during RGB evolution). This mass was selected to
have in RGB an age as close as possible to the mean estimated age value
for GCs, i.e. about 12 Gyr.  The progenitor masses satisfying this
constraint are reported in Table~\ref{table:mass-zahb}; in the most cases the variation of $\alpha$ enhancement and mixing length
values does not affect the progenitor mass determination. As is well
known, the small dependence of HB characteristics on cluster age can be
neglected; a detailed investigation of the age effect on HB models can
be found e.g. in \citet{caputo2002}.  
A total of 16081 models are included in this part of the database.

Lower main-sequence stars ignite helium in a violent flash at the RGB
tip; following a common procedure \citep{dorman1991, castellani1989},
instead of modeling this phase, we stopped our calculations of the ZAHB
progenitor at the RGB flash,
defined as the time when the He
  burning luminosity reaches 100 times the surface luminosity.  The He
  core mass at this time was assumed as the core mass of the starting models of
  quiescent central helium burning.  In all cases an
  initial amount of carbon, given by X$_{\rm C}$ = 0.03, was assumed to
  be homogeneously distributed throughout the He core, as a product of
  the He burning during the flash. The chemical composition of the
  models out to the He core was taken as the external one at the He
  flash; in this way we also took into account the external helium
  overabundance with respect to the MS (extra-helium) driven to the
  surface by the first dredge-up, which also homogenizes the stellar
  chemical composition out of the He core.  Model calculations were therefore
  started again as thermal relaxed models in the central He
  burning phase; ZAHB point was fixed when the equilibrium
  abundance of CNO burning secondary elements was reached, after about
  1 Myr.  The mass of the H-rich envelope was taken as a free
  parameter, in dependence on the unknown amount of mass loss
  experienced in the RGB phase by real cluster stars. In practice,
  several HB models with fixed He-core mass and external
  chemical abundance, but different total masses were computed in
  a way to homogeneously cover the ZAHB extension in effective temperature.  
  We started
  by creating the first He burning model corresponding to the bluest
  one. The new model was found by a Runge-Kutta integration (more
  precisely the ``fitting method'', as described in \citealt{Kippenhahn}).
  After creating this model for the
  lowest mass star, higher mass He burning models, up to one
  corresponding to the progenitor mass, were calculated by increasing
  the envelope mass to the required values. During this procedure time
  steps were artificially kept very short to prevent model evolution.

Comparisons among fully evolved models and
horizontal branch models constructed in the described way confirm the
reliability of this ZAHB model building procedure 
\citep[see e.g.][]{serenelli2005, piersanti2004, vandenBerg2000}.

\begin{table*}
\caption{Summary of RGB progenitor mass (in $M_{\sun}$) for the computed HB
  grids.} 
\label{table:mass-zahb} 
\centering          
\begin{tabular}{c|c|c|c|c|c}    
\hline\hline  
\multicolumn{6}{l}{Mixture: \citet{AGSS09}, [$\alpha$/Fe] = 0.0, 0.3}\\
\multicolumn{6}{l}{$\alpha_{ml}$ = 1.70, 1.80, 1.90}\\
\hline
$Z$ & \multicolumn{5}{c}{$Y$}\\
   & 0.25 & 0.27 & 0.33       & 0.38 & 0.42\\
\hline 
0.0001 & 0.80 & 0.75 & 0.70       & 0.60 & 0.55\\
0.0002 & 0.80 &	0.75 & 0.70       & 0.60 & 0.55\\
0.0003 & 0.80 &	0.75 & 0.70       & 0.65 & 0.60\\
0.0004 & 0.80 &	0.75 & 0.70       & 0.65 & 0.60\\
0.0005 & 0.80 &	0.75 & 0.70       & 0.65 & 0.60\\
0.0006 & 0.80 &	0.75 & 0.70       & 0.65 & 0.60\\
0.0007 & 0.80 &	0.80 & 0.70       & 0.65 & 0.60\\
0.0008 & 0.80 &	0.80 & 0.70       & 0.65 & 0.60\\
0.0009 & 0.80 &	0.80 & 0.70       & 0.65 & 0.60\\
0.001  & 0.80 &	0.80 & 0.70       & 0.65 & 0.60\\
0.002  & 0.85 &	0.80 & 0.70       & 0.65 & 0.60\\
0.003  & 0.85 &	0.85\tablefootmark{***} &	0.75 &	0.65 &	0.60\\
0.004  & 0.90\tablefootmark{*} & 0.85 &	0.75 &	0.70 &	0.65\\
0.005  & 0.90 &	0.85 & 0.75       & 0.70 & 0.65\\
0.006  & 0.90 &	0.90 & 0.80       & 0.70 & 0.65\\
0.007  & 0.90\tablefootmark{**} &	0.90 &	0.80 &	0.70 & 0.65\\
0.008  & 0.95 &	0.90 &	0.80      & 0.75 & 0.65\\
0.009  & 0.95 &	0.90 &	0.80      & 0.75 & 0.70\\
0.01   & 0.95 &	0.95 &	0.85      & 0.75 & 0.70\\
\hline 
\end{tabular}
\tablefoot{\\
\tablefoottext{*}{0.85 for $\alpha_{ml}$ = 1.90, [$\alpha$/Fe] = 0.3}\\
\tablefoottext{**}{0.95 for $\alpha_{ml}$ = 1.70, [$\alpha$/Fe] = 0.0}\\
\tablefoottext{***}{0.80 for $\alpha_{ml}$ = 1.90, [$\alpha$/Fe] = 0.0, 0.3}\\
}
\end{table*}

The ZAHB total masses were chosen to span a sizeable range of the ZAHB 
effective temperature extension: from zero mass loss in RGB (ZAHB
mass equal to the progenitor mass) to a mass equal to that of the He core 
at RGB flash plus a small envelope of 0.026 $M_{\sun}$.
The ZAHB models were calculated in intervals of 0.01
$M_{\sun}$ in mass to avoid spurious discontinuities in the ZAHB morphology. 
Each ZAHB 
table contains mass in $M_{\sun}$, effective temperature in $\log T_{eff}$
and luminosity in $\log L/L_{\sun}$. The file names were chosen in the same
way as the track 
files:
 as an example, for $M$ = 0.80 $M_{\sun}$,
$Z$ = 0.001, $Y$ = 0.25, $\alpha_{ml}$ = 1.90, solar-scaled Asplund 2009
 mixture, 
the ZAHB file is named {\it
  ZAHB\_M0.80\_Z0.00100\_He0.2500\_ML1.90\_AS09a0.DAT}. 

Isochrones are stored in several directories with self 
explicative names; as an example the
directory {\it ISO\_Z0.00100\_He0.2500\_ML1.90\_AS09a0.DAT}
contains all isochrones with the indicated chemical
composition and convection efficiency.
The directory hosts several files for the different ages. As an example for a
8.0 Gyr isochrone the file is named
 {\it
  AGE08000\_Z0.00100\_He0.2500\_ML1.90\_AS09a0.DAT}.
The header of these files lists the age in Gyr, 
the $Y$ and $Z$ content, the adopted value for $\alpha_{ml}$ and the solar
mixture. The possible $\alpha$ enhancement is specified both in the file name
and in the header. 
For each isochrone the reported quantities are
the luminosity in $\log L/L_{\sun}$, the effective temperature in $\log
T_{eff}$ and the mass of the star in $M/M_{\sun}$.

Owing to the extremely wide range of possible useful photometric bands and to
the dependence of the obtained colors (mainly for cool models) on the adopted
color 
transformations we decided to present results in the theoretical plane only,
delaying the presentation of our calculations in several observational planes
to a 
following paper, in which the different sources of uncertainties for theoretical
evolutionary models will also be discussed.

\begin{figure*}
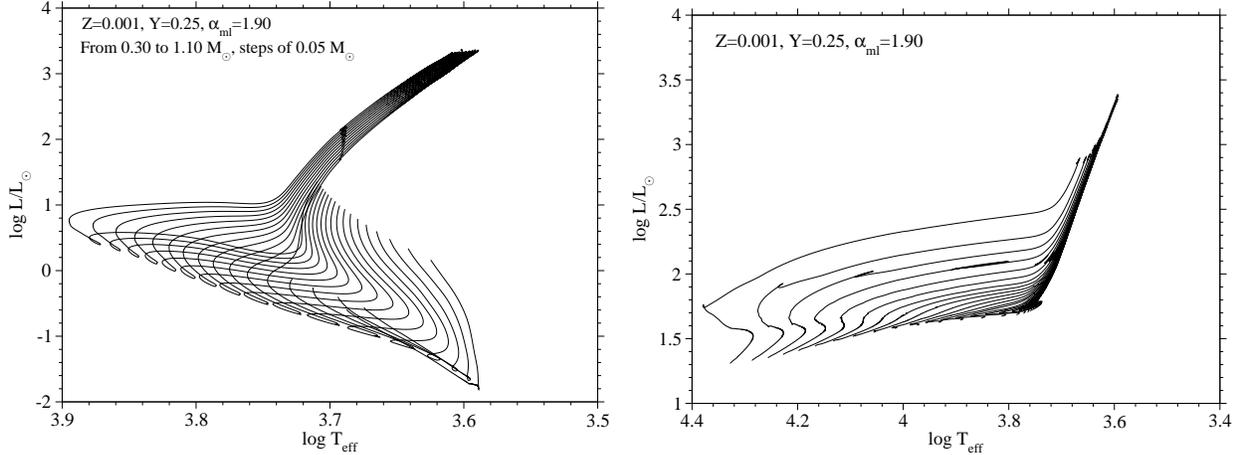

\centering
\includegraphics[width=8cm]{tracce-z0.001.eps}\hspace{2mm} 
\includegraphics[width=8cm]{ZAHBgrid0.001.eps}
\caption{HR diagram for evolutionary tracks in the mass range 0.30 $\div$ 1.10
$M_{\sun}$ for the labeled chemical composition ([$\alpha$/Fe] = 0.0)
and $\alpha_{ml}=1.90$. Effective temperatures are in K. Left panel: tracks
from PMS up to the central 
hydrogen exhaustion (for masses up to 0.5 $M_{\sun}$) and tracks from PMS up
to the helium flash (for masses $0.55 \; M_{\sun} \leq M \leq 1.10 \;
M_{\sun}$). Right panel: HB models with a 0.80 
$M_{\sun}$ progenitor, from the ZAHB to thermal pulses.  }
 \label{fig:tracce}
\end{figure*}

\begin{figure*}
\centering
\includegraphics[width=8cm]{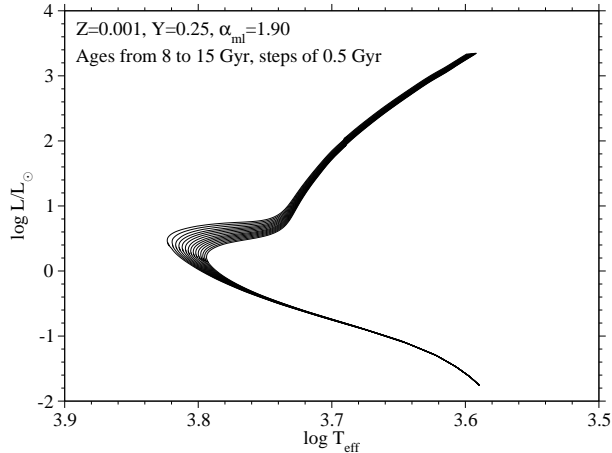}
\caption{Theoretical isochrones in the range $8 \div 15$ Gyr for the labeled
chemical compositions with $\alpha_{ml}$ = 1.90 and [$\alpha$/Fe] = 0.0.}
\label{fig:isocrone}
\end{figure*}

Examples of present calculations are shown in Figs.~\ref{fig:tracce} and 
\ref{fig:isocrone},
where tracks and isochrones for $Z=0.001$, $Y=0.25$, $\alpha_{ml}=1.90$
are plotted in the ($\log L/L_{\sun}$, $\log T_{eff}$) plane.

\section{Comparison with recent stellar model
  databases}\label{sec:db-comparison} 

In this section present results for some of the most relevant
evolutionary parameters (namely the TO and RGB tip, TRGB, luminosity and the
He core mass at the RGB tip, $M_c^{He}$)
are compared with those of other recent papers available in
the literature. As is well known, the TO luminosity is the most important age
indicator for globular clusters while tip and horizontal branch luminosity
are powerful distance indicators. 
The TRGB and HB luminosities are  proportional to $M_c^{He}$
 \citep[see e.g.][]{salaris1998, buzzoni1983}, which also affects
RGB and HB lifetimes.  

We selected works that provide an extended database of tracks and isochrones 
with solar-scaled and $\alpha$-enhanced chemical compositions for the required age
range: BaSTI database \citep{teramo04,teramo06}, Dartmouth
\citep{dotter07,dotter08},  
and Padova STEV \citep{padova08,padova09}. Obviously, the
quoted models do not exhaust the rich and composite
scenario of stellar evolutionary calculations. Updated stellar models differ
in a  variety of choices concerning
physical inputs and chemical compositions, which produces small but significant
changes 
in the results. Models with physical inputs that differ within the
error of each quantity are all acceptable and are useful
for estimating the order of magnitude of the uncertainties on the theoretical
predictions. 
The main differences in the inputs of the various codes used by the different authors are
summarized in Table \ref{table:2}. From top to bottom we show 
mass range, metallicity and helium values, the adopted
mixture, the $\alpha$ enhancement and the evolutionary phases
followed by the models; then we report the adopted mixing length
parameter, and the presence or absence of diffusion and overshooting.  The
following lines list the adopted physical inputs: equation of
state (EOS), radiative opacity, nuclear reaction rates, electron conduction in
degenerate matter, neutrino energy loss rates. Information about the
selected atmospheric model are reported on the last line.

For completeness we notice that our models are suitable for old
clusters and therefore they are restricted to lower main-sequence stars which,
except for masses of $\approx 1.1$ $M_{\sun}$,
have radiative cores; the other quoted databases cover a more extended range
of masses including also upper main-sequence stars. Overshooting influences
only stars with a mass greater than about 1.1 $M_{\sun}$ (depending on the
chemical composition), which start developing small convective cores. Its
efficiency is usually modeled to grow linearly with mass, until about 1.5 $\div$
1.7 $M_{\sun}$. In addition to models computed without
overshooting, there are also models in the BaSTI
database that adopt a gradual
increase of the overshooting efficiency (usually expressed in units of
pressure scale height) $\Lambda_c = (M/M_{\sun}-0.9)/4$
\citep{teramo04} for $1.1 \leq M < 1.7 \;
M_{\sun}$. The Padova database adopts
$\Lambda_c = M/M_{\sun} - 1.0$ \citep{padova08} for $1.0 < M < 1.5M_{\sun}$. In the Dartmouth database
the convective core overshooting is linearly increased from 5\% of
the pressure scale height for $M = M_{min}$ (the value of $M_{min}$ depends
on chemical composition and is given in Table 3 of \citet{dotter07}), to
20\% for $M = M_{min} + 0.2 \; M_{\sun}$.  Above these limits the overshooting
efficiency is assumed to be constant. Thus the inclusion of overshooting in
the models could slightly influence our comparison only for masses of about 
1.1 $M_{\sun}$; in Table \ref{table:2} we report for each database only
the minimum mass in which convective core overshooting is included.

Each database, except the STEV, includes for the chemical composition,
models with helium abundance calculated with the quoted
linear relation between helium and metal enrichment with a primordial helium
abundance of $Y=0.245$ and a relation coefficient of $\approx 1.5 \div
2$. Moreover, each database spans a wide range of metallicities and helium
abundances; this enabled us to select for our comparison the most similar
chemical compositions among those available.

\begin{table*}
\centering
\caption{Comparison among recent databases.}             
\label{table:2}      
\centering
\begin{tabular}{l l l l l}     
\hline\hline       
Models              &   
Present models (Pisa)    &  
BaSTI (Teramo)      & 
STEV(Padova)        & 
Dartmouth \\
\hline                    
Mass range [$M_{\sun}$] &
$0.30\div1.10$        &
$0.50\div10.0$        &
$0.15\div20.0$        &
$0.10\div4.00$  \\
Metallicity ($Z$)                           &
$1 \times 10^{-4}\div 1 \times 10^{-2} $    &
$1 \times 10^{-4}\div 4 \times 10^{-2} $    &
$1 \times 10^{-4}\div 7 \times 10^{-2} $    & 
[Fe/H] = $-2.5\div0.5$ \\
Helium abundance ($Y$)                      &
0.25; 0.27; 0.33; 0.38; 0.42;   &
0.245$\le Y \le$0.303           &
0.23; 0.26; 0.30; 0.34;         &
0.33; 0.40; \\
                                &
$Y=0.2485+2Z$                   &
$Y=0.245+1.4Z$                  &
0.40; 0.46  \tablefootmark{*}                    & 
$Y=0.245+1.6Z$ \\
Solar mixture                         &
AGSS09\tablefootmark{a}         &
GN93\tablefootmark{b}           &
GN93\tablefootmark{b}           &
GS98\tablefootmark{c}\\
${\rm [}\alpha{\rm/Fe]}$                    &
0.0; +0.3                       &
+0.4                            &
0.0                             &
$-0.2\div+0.8$ ([Fe/H] $\le 0$)\\
                                & 
                                & 
                                & 
                                &
$-0.2\div+0.2$ ([Fe/H] $>0$)\tablefootmark{**}\\
Evolutionary phases           & 
PMS; H+He              &
PMS; H+He              &
H+He                   &
PMS; H+He \\
\hline 
$\alpha_{ml}$  &1.70 ; 1.80 ; 1.90  & 1.913     &1.68 & 1.938 \\
Diffusion &\citet{thoul94}  &NO     &NO &\citet{thoul94}  \\
$M_{min}$ for overshoot& NO overshooting &1.1 $M_{\sun}$ &1.1 $M_{\sun}$  &1.1 $M_{\sun}$\\
\hline
EOS&OPAL2006+\citet{Straniero88} & FreeEOS\tablefootmark{A} &\citet{padova08}+ & \citet{chaboyerkim95}+\\
&&& \citet{MHD1990}&FreeEOS\tablefootmark{A}  \\
Radiative opacity&OPAL2006+F05\tablefootmark{d}
&OPAL96+F05\tablefootmark{d}&OPAL96+AF94\tablefootmark{e}
&OPAL96+F05\tablefootmark{d}\\
Conductive opacity&\citet{casspote07} &\citet{pote99a}&\citet{itoh83}
&\citet{hubbard69}\\
&&\citet{pote99b}&&\\ 
Reactions rates&NACRE   &NACRE  &\citet{caughlan88} &\citet{adel98}\\
              &\citet{14n}\tablefootmark{f}&
& &\citet{Imbriani2004}\tablefootmark{f}\\
              &\citet{12c}\tablefootmark{g} &\citet{kunz02}\tablefootmark{g}&
&\citet{kunz02}\tablefootmark{g}\\ 
              &\citet{cyburt08}\tablefootmark{h}&
&\citet{landre90}\tablefootmark{i} &\\ 
Neutrinos&\citet{haft94} &\citet{haft94} &\citet{haft94}
&\citet{haft94}\\
&\citet{itoh96} & & &\\
\hline  
Boundary conditions  &\citet{brott05} & \citet{KrishnaSwamy1966} &\citet{castelli03}
&\citet{hauschildt99} \\
&\citet{castelli03} &  && \citet{castelli03}\\
\hline 
\end{tabular}
\tablefoot{\\
\tablefoottext{*}{not all values of $Z$ are available for all $Y$ values}\\
\tablefoottext{**}{not all values of [Fe/H] are available for the reported
  [$\alpha$/Fe] values}\\ 
\tablefoottext{A}{http://freeeos.sourceforge.net}\\
\tablefoottext{a}{AGSS09=\citet{AGSS09}},
\tablefoottext{b}{GN93=\citet{GN93}},
\tablefoottext{c}{GS98=\citet{GS98}}\\
\tablefoottext{d}{F05=\citet{ferg05}},
\tablefoottext{e}{AF94=\citet{ferg94}}\\
\tablefoottext{f}{for the reaction rate $^{14}N(p,\gamma)^{15}O$}\\
\tablefoottext{g}{for the reaction rate $^{12}C(\alpha,\gamma)^{16}O$}\\
\tablefoottext{h}{for the reaction rate $^{3}He(\alpha,\gamma)^{7}Be$}\\
\tablefoottext{i}{for the reaction rate $^{17}O(p,\gamma)^{18}F$}\\
}
\end{table*}

\begin{figure*}
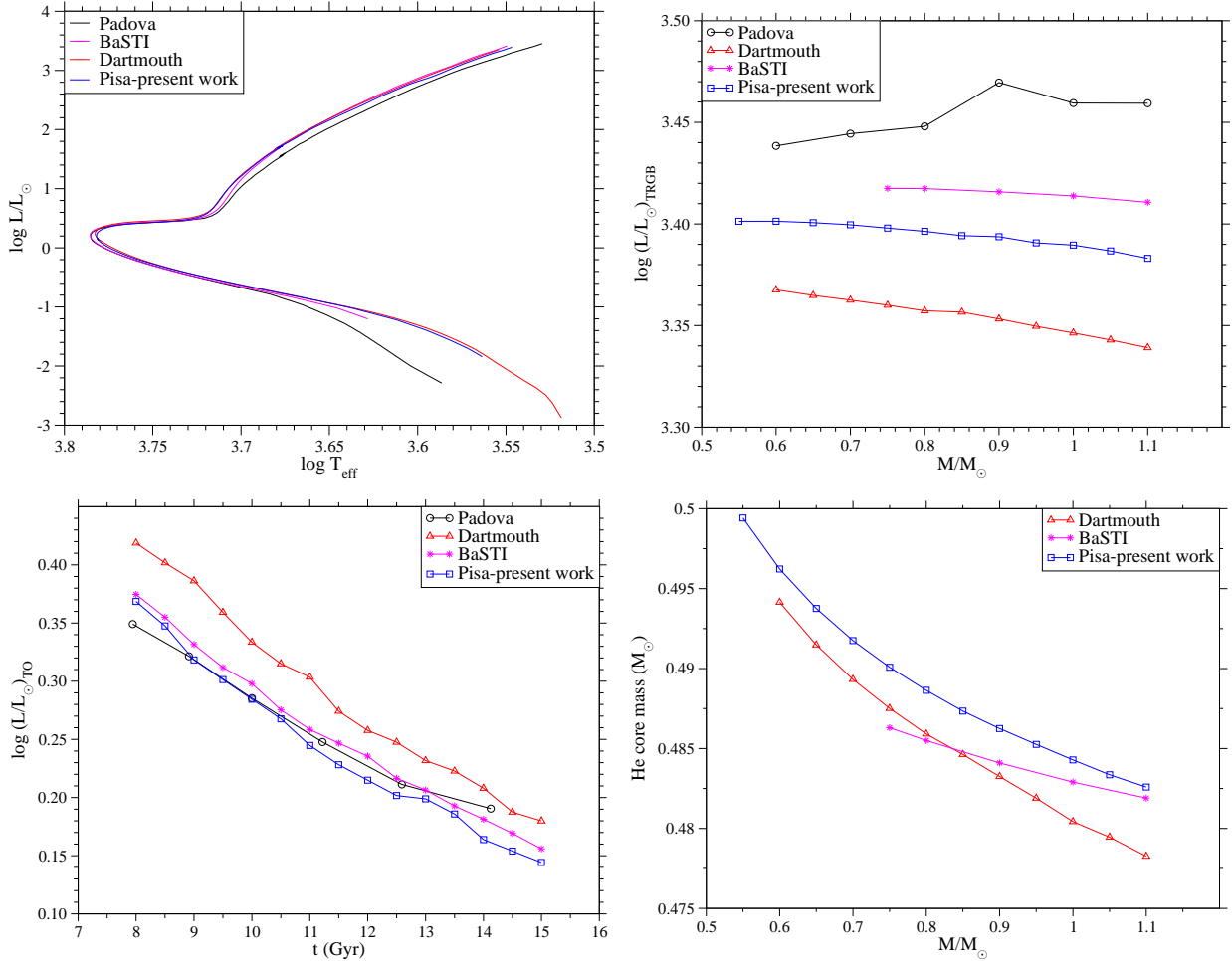

\centering
\includegraphics[width=8cm]{cfr-age_12.5_z0.0040_y0.251.eps} 
\hspace{2mm}        
\includegraphics[width=8cm]{cfr-lTIP.eps}\\\vspace{2mm}
\includegraphics[width=8cm]{cfr-lto.eps}\hspace{2mm}
\includegraphics[width=8cm]{cfr-mcHe.eps}
\caption{Comparison at $Z=0.004$, $Y=0.25$ and $\alpha_{ml}$ = 1.90 among the
  different databases of Table \ref{table:2}. For the STEV database, we
  selected $Y=0.26$ and $\alpha_{ml}$ = 1.68 as the values among those
  available that are
  closest to those of the other databases. The tracks of the Dartmouth
  databases were interpolated in $Z$, see text.  Upper left panel:
  theoretical isochrones at $t$ = 12.5 Gyr. Upper right panel: luminosity at the
  tip of the red giant branch. Lower left panel: turn-off luminosity. Lower right
  panel: mass of the helium core at the He flash.}
\label{fig:confronto-iso}
\end{figure*}

\begin{figure*}
\centering
\includegraphics[width=8cm]{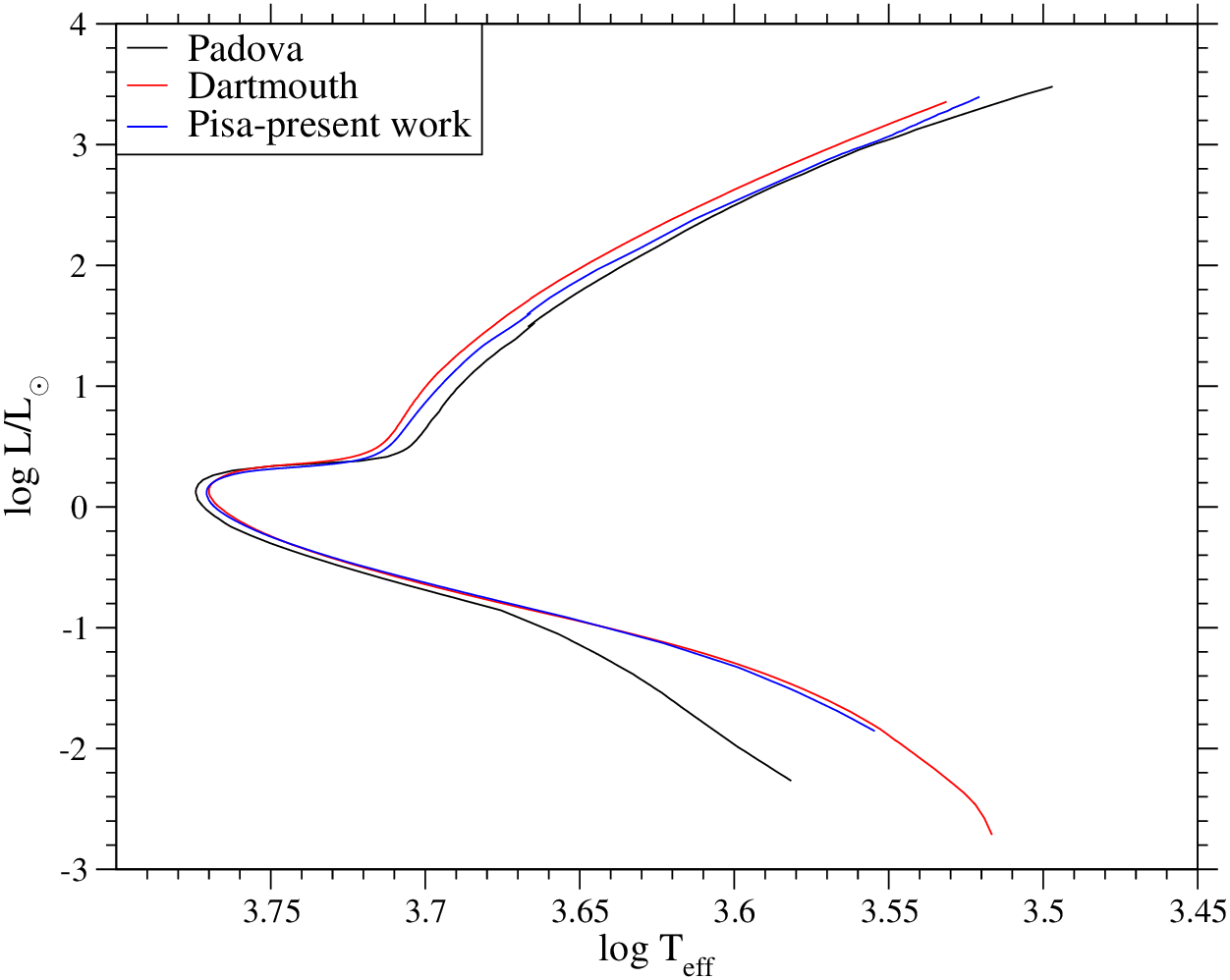}\hspace{2mm} 
\includegraphics[width=8cm]{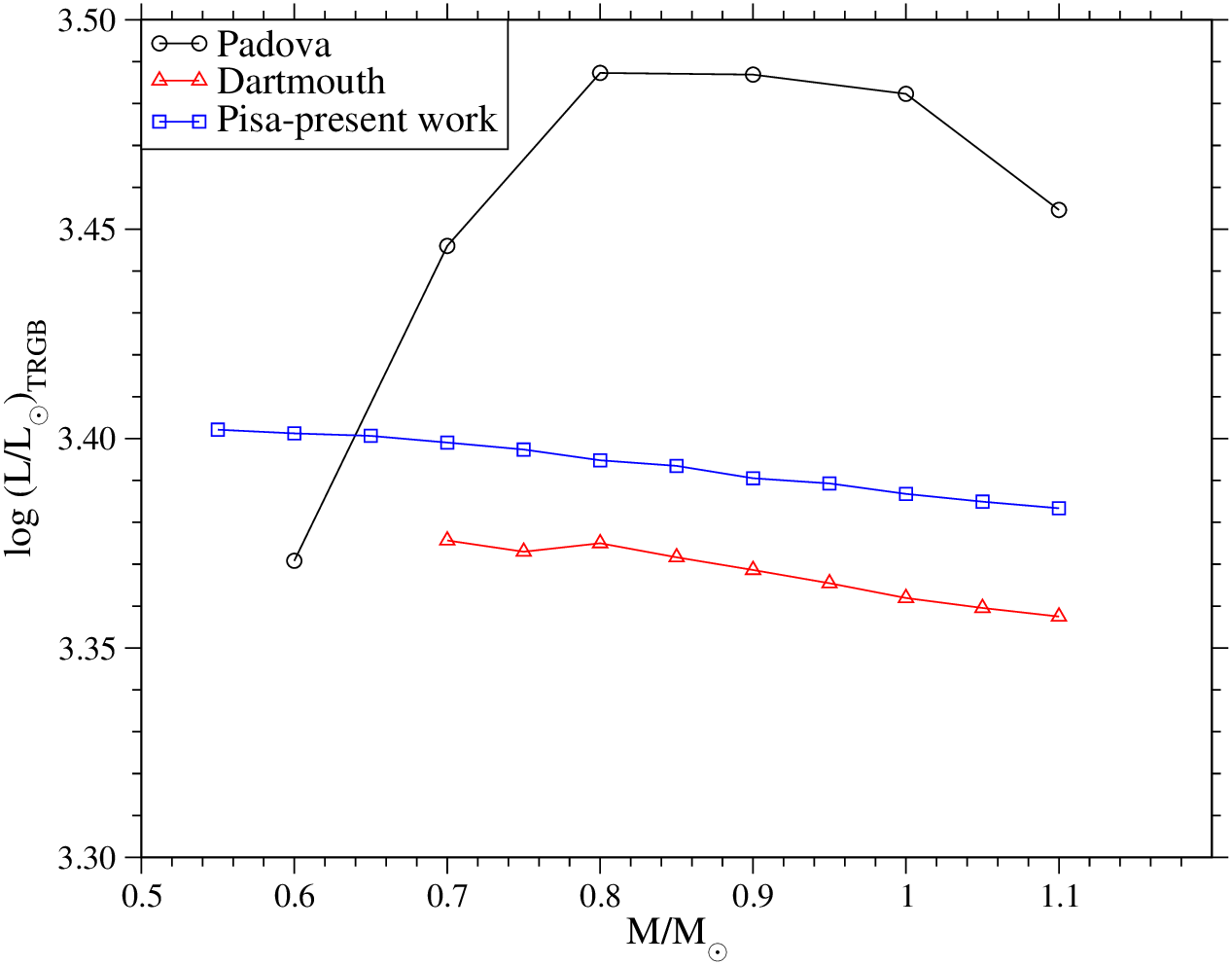}\\\vspace{2mm}
\includegraphics[width=8cm]{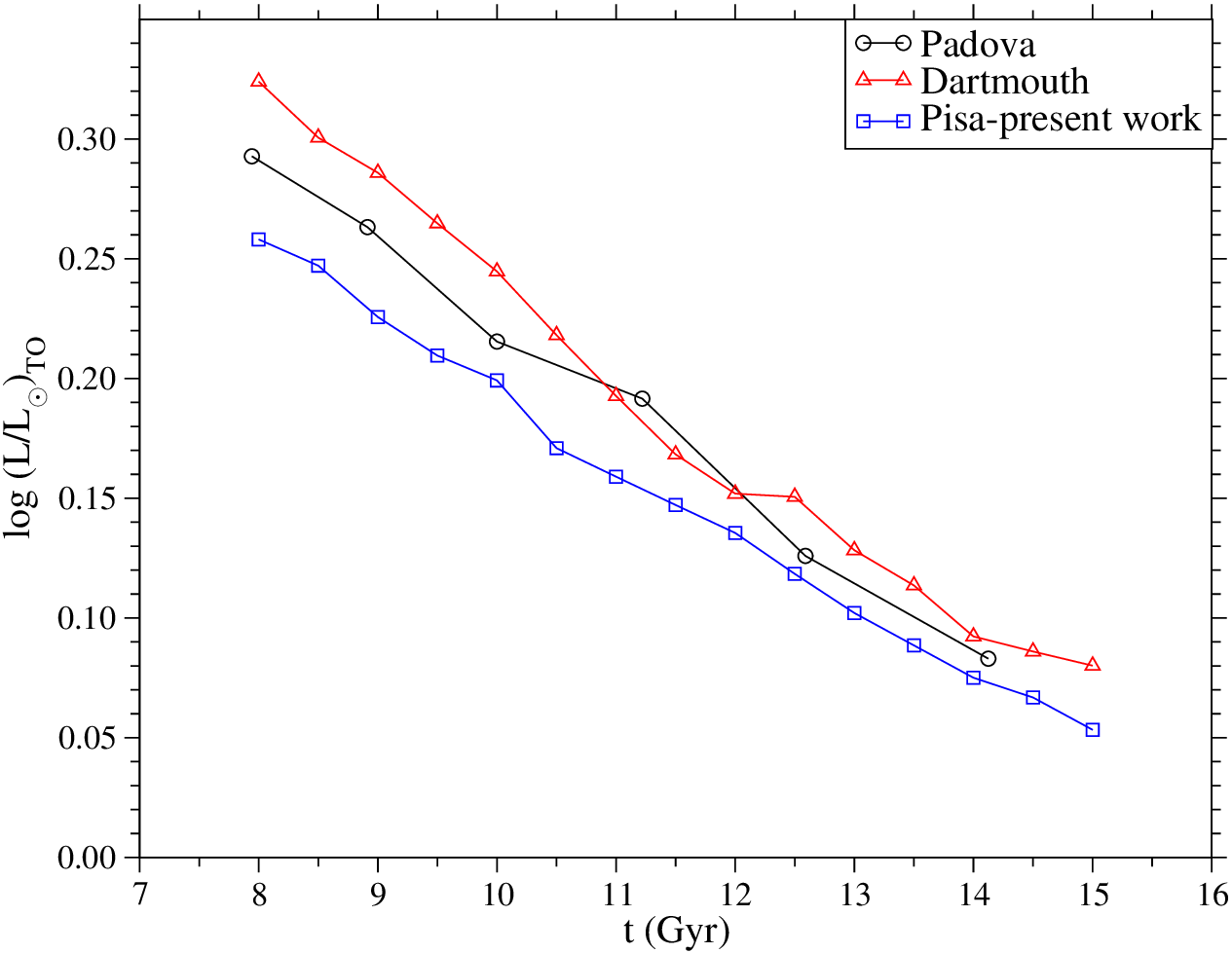}\hspace{2mm}
\includegraphics[width=8cm]{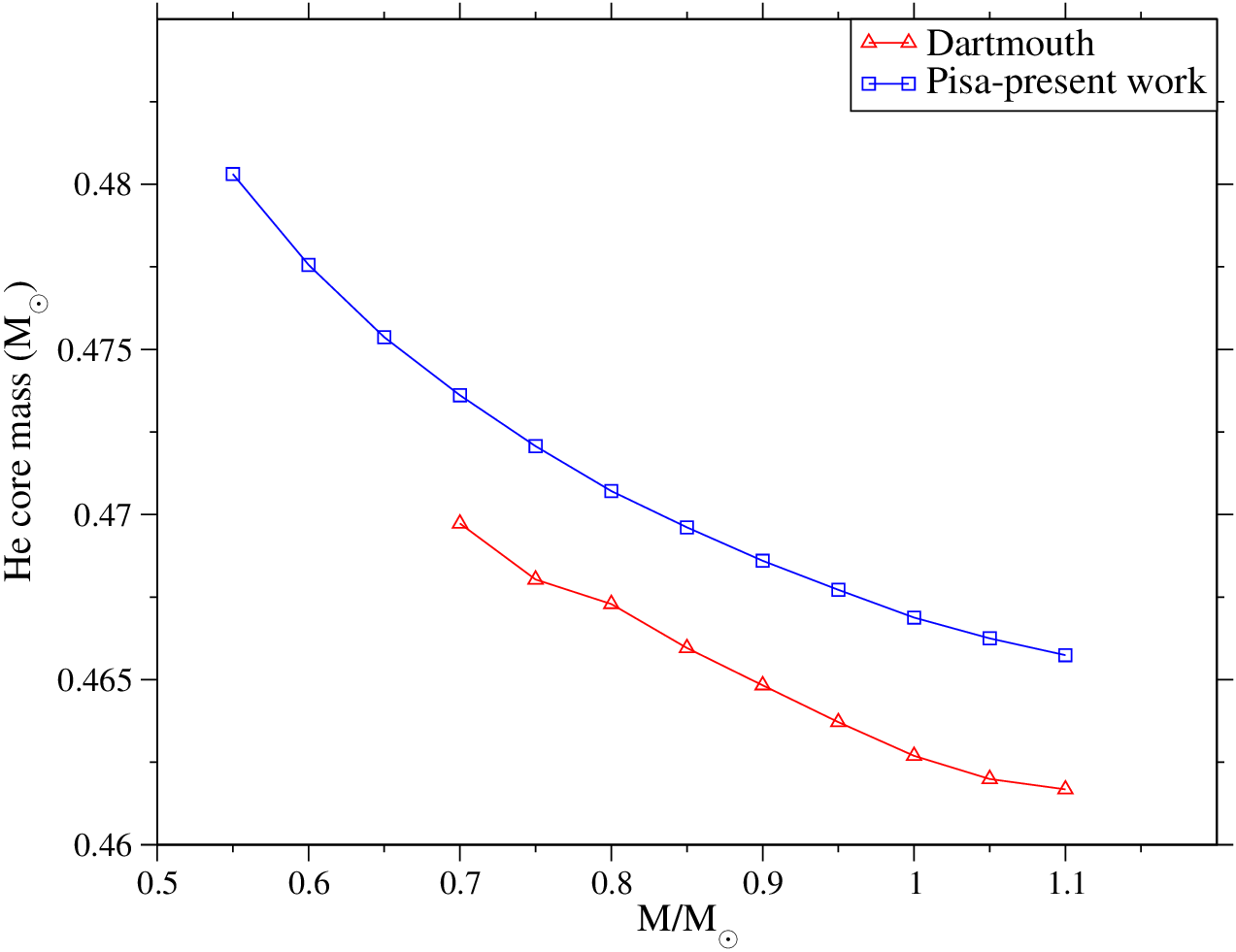}
\caption{Comparison at $Z=0.008$, $Y=0.33$ and $\alpha_{ml}$ = 1.90 among the
  different databases of Tab.\ref{table:2}. For the STEV database, we
  selected $Y=0.34$ and $\alpha_{ml}$ = 1.68 as the values among those
  available that are
  closest to those of the other databases. The tracks of the Dartmouth
  databases were interpolated in $Z$, see text.  Upper left panel:
  theoretical isochrones at $t$ = 12.5 Gyr. Upper right panel: red giant
  branch tip luminosity. Lower left panel: turn-off luminosity. Lower right
  panel: Helium core mass at the He flash.}
\label{fig:confronto-iso-y033}
\end{figure*}

We present two comparisons. The first one, in Fig.~\ref{fig:confronto-iso},
 with $Z=0.004$ and $Y=0.25$,
except for the STEV database for which 
an helium value $Y=0.26$ is available; the effect of this helium
variation on the analyzed evolutionary parameters is known to be very
small \citep[see e.g.][]{buzzoni1983}. The second one, in
Fig.~\ref{fig:confronto-iso-y033}, with  
$Z=0.008$ and $Y=0.33$,
except for the STEV database for which $Y=0.34$ is available; 
in this case we were unable to select a
model from BaSTI with the required $Z$ abundance.
The value of $Z=0.004$ and $Z=0.008$ were
unavailable in the Dartmouth databases: the isochrones (upper panels in 
Figs.~\ref{fig:confronto-iso} and \ref{fig:confronto-iso-y033}) 
were interpolated in
$Z$ with a cubic interpolator available on their web
site\footnote{\url{http://stellar.dartmouth.edu/~models/isolf.html}}, while the
evolutionary quantities (lower panels in 
Figs.~\ref{fig:confronto-iso} and \ref{fig:confronto-iso-y033}) were
interpolated in $Z$ by us with a linear interpolator. 

Moreover, there are differences in the
solar mixture adopted by the different databases. Recent
analysis of spectroscopic data using three dimensional hydrodynamic
atmospheric models \citep[see][]{asplund05, AGSS09} have reduced the derived
abundances of CNO and other heavy elements with respect to the previous
estimate by \citet{GS98} (hereafter GS98), even if additional investigations are
needed  
\citep[see e.g.][]{caffau2009, SocasNavarro2007}. If one takes into
account the still widely used solar mixture by \citet{GN93}, with C, N and O
abundances slightly higher than those by GS98, the discrepancy with the
\citet{asplund05, AGSS09} composition slightly increases. 
It is worth noticing that uncertainties on the solar mixture have two main 
effects: a variation of the relation between [Fe/H] and total metallicity $Z$,
and a change of the model characteristics at fixed $Z$. 
For the present comparison
at fixed $Z$, we are interested in the second point; fortunately, it is
already demonstrated \citep{DeglInnocenti2006} 
that the influence of the adopted mixture on model
luminosities and He core mass is very small,
while effective temperatures could somehow be affected 
\citep[see e.g.][]{salaris1993}. 

Figs.~\ref{fig:confronto-iso} and \ref{fig:confronto-iso-y033} 
show the results of the comparison among the chosen databases
for the isochrone HR diagrams and for the selected evolutionary 
quantities. For the isochrone comparison we selected the age of 12.5 Gyr
because this is a value common to all the databases.

As one can see in Table \ref{table:2}, the various databases are
computed adopting different choices of the physical inputs; the source of
the opacities is quite often the same (except for STEV, which adopts values for low-temperature opacities from \citealt{ferg94}) but the EOS, nuclear
reaction rates, boundary conditions and electronic conduction 
are often different. Moreover, two of the selected
databases (Pisa and Dartmouth) are calculated including microscopic diffusion
and helium and 
heavy elements (with the same diffusion coefficients), while the other two
databases neglect the diffusion process.

Consequently, a precise quantitative analysis of the differences in the results among
the various databases would require the ``ad hoc'' calculation of several
models with different physical inputs, which is beyond the scope of this
paper. However, as discussed before, even a more qualitative analysis of the
differences is useful to give an indication of the still present uncertainties
due to the adoption of different physical inputs in stellar codes.

We did not perform a comparison among the different horizontal branch models
because  
the other databases do not make the corresponding mass grids available. 

Elements diffusion occurs on a timescale of a few Gyr, so it influences the
main physical characteristics of old clusters only.  Diffusion has also been
demonstrated to be efficient in the Sun \citep[see e.g.][]{Bahcall2001, 
Guzik2001} for which the huge amount of very precise observational data allow
one to observe 
effects smaller than those occurring in old clusters.
The general quoted uncertainty on this process, \citep[see
e.g.][]{thoul94}, is on the order of 10$\div$15\%; however, the treatment of
microscopic diffusion still presents several uncertainties even for the Sun
for which a very large set of observational data is available 
\citep[see e.g.][]{Thoul2007, Montalban2006, Richer2000, Turcotte1998}.
Surface abundance observations in globular clusters have raised some doubts
about the actual  
efficiency of microscopic diffusion in old cluster stars \citep[see
e.g.][]{James2004, Gratton2001}, but these results seem to not be confirmed by
recent analysis \citep[see][and references therein]{Korn2007, Lind2008}.

A detailed discussion of the influence of the helium and metal microscopic
diffusion on evolutionary properties can be found in \citet{castellani1999}
\citep[see also][]{Straniero1997, Castellani1997}. 
Here we only recall that including microscopic diffusion at a fixed age
reduces the TO luminosity by $\Delta \log L/L_{\sun} \approx 0.06$; 
moreover, He is ignited within a slightly larger He core with a lower He
abundance in the envelope, so that the ZAHB luminosity (in the RR Lyrae region
at 
about $\log L/L_{\sun} =3.83$) is slightly decreased, while the TRGB
luminosity is 
almost unaffected \citep[see also][]{Cassisi1998}. The results is that
neglecting diffusion leads to an increase of the estimated age through the
``vertical method'' \citep{Iben1968} by $\approx$ 1 Gyr.

Another important point is that only two databases adopt the recent value of
the $^{14}$N(p,$\gamma$)$^{15}$O astrophysical factor by the LUNA
Collaboration \citep[][and references therein]{14n},
which is about half
of the previous quoted estimates.  Some authors 
\citep[see e.g.][]{Weiss2005, Imbriani2004, DeglInnocenti2004} 
analyzed the effects
of this cross section update showing that TO luminosity is increased by about
0.03 in $\log L/L_{\sun}$ while the influence on HB luminosity in the RR Lyrae
region is a factor three smaller and it also depends on model metallicity;
moreover, \citet{Pietrinferni2010} showed that using the LUNA cross section
causes an increase of the $M_c^{He}$ by about 0.002-0.003 $M_{\sun}$ and a
decrease of the TRGB luminosity (because of the lower CNO burning efficiency) of
$\Delta \log L/L_{\sun} \approx$ 0.01-0.02 dex.

Selected databases adopt also different, and sometimes not updated,
evaluations for conductive opacities that significantly affect
the He core mass at the helium ignition, and consequently the HB luminosity
\citep[see 
e.g.][]{castellani1999, Catelan1996}. \citet{castellani1999} noticed 
that the adoption of the \citet{itoh83}
evaluations, present in the STEV database, instead of the \citet{hubbard69} 
ones, adopted by the Dartmouth database, leads to an increase of the
helium core of about 0.005 $M_{\sun}$ and thus to a corresponding increase of
the ZAHB luminosity by $\Delta \log L/L_{\sun} \approx 0.017$. 
Moreover, \citet{casspote07} pointed out that the adoption of the 
\citet{pote99b}
conduction opacities provides $M_c^{He}$ values between those obtained with
the \citet{itoh83} and \citet{hubbard69} ones, but closer to the
\citet{itoh83} results.  
In  \citet{casspote07} the opacity calculations by \citet{pote99a} and
\citet{pote99b}  
were improved by including the
electron-electron scattering in partially degenerate and non degenerate
matter. The authors found that the change of the conduction treatment from
\citet{pote99b} to \citet{casspote07} leads to a reduction of
$M_c^{He}$ by about 0.006 $M_{\sun}$ with a corresponding decrease of TRGB
luminosity of $\Delta \log L/L_{\sun} \approx 0.03$ and of the ZAHB luminosity
by $\Delta \log L/L_{\sun} \approx 0.02$.

From Table \ref{table:2} one sees that the models belonging to different
databases adopt different EOS,  hydrogen and helium
burning nuclear reaction rates. The result 
is that the differences among the predicted evolutionary quantities are due to a
combination of the effects of all the quoted physical input variations in a
way that is difficult to disentangle.  

For example, the slightly lower TO luminosity of the present models with respect
to the BaSTI ones (see Fig.~\ref{fig:confronto-iso}) can be understood 
in terms of the effects of the diffusion
inclusion and of the $^{14}$N(p,$\gamma$)$^{15}$ update, taking also into
account that the adopted EOS is quite similar and that both databases take
most of the reaction rates from  NACRE compilation. However, 
even if present models
and Dartmouth calculations adopt the same microscopic diffusion coefficients  
and $^{14}$N(p,$\gamma$)$^{15}$ cross section, TO luminosity differs up to 
$\Delta \log L/L_{\sun} \approx 0.067$; which is probably at least in part
due to the different EOS and H burning reaction rates.

From the isochrone comparison of Figs.~\ref{fig:confronto-iso} and 
\ref{fig:confronto-iso-y033}, differences in effective temperature
appear evident mainly in RGB and in the lower main-sequence.
Differences in the RGB location among  the various databases are not
surprising because 
its effective temperature is very sensitive to low-temperature opacities, 
external convection efficiency, and outer boundary
conditions. Particularly, the effective temperature of the upper part of the 
RGB (at luminosity higher than
the RGB bump) of the STEV isochrone differs from the others.
 
There is a fair agreement among the $M_c^{He}$ values from the different
databases that make this quantity available; the maximum difference is on
the order of 0.005 $M_{\sun}$, 
fully compatible with the adoption of the different physical inputs described 
above.

All the models, except the STEV, agree
within $\Delta \log L/L_{\sun} \approx 0.06$ for the TRGB luminosity; even if it is obvious from the
$M_c^{He}$ behavior that He core mass is not the only parameter that
influences the luminosity at the He flash.

\section{Analytical relations}\label{sec:modelfit}

The wide range of chemical compositions spanned by our database and its
fine spacing in the input parameters is particularly suitable for the 
calculation of
analytical relations, which express the dependence of the main evolutionary
characteristics on the various parameters allowing one to identify the
critical input factors for each selected evolutionary feature.  Moreover,
 sufficiently precise enough relations allow one to obtain the required evolutionary results
also for a combination of parameters for which models are not directly
calculated and can be useful for comparison with other theoretical
predictions. 

Analytical relations connecting relevant evolutionary quantities with
stellar masses and ages can be useful in several fields of stellar
evolution, e.g. evolutionary properties of binary systems, synthetic
models for simple stellar populations and for star counts in galaxies,
chemical evolution models of galaxies  \citep[see e.g.][for representative
works in the quoted fields]{Andersen2002, Andersen1991, Bahcall1980, 
Chiappini1997, Bruzual2003, Bruzual1993, Haywood1994, Matteucci2009, 
Popper1997, Hernandez2000, Portinari2000, 
Ribas2000a, Ribas2000b}.

We analyzed two relevant evolutionary features: the TO luminosity
and the 
ZAHB luminosity in the RR Lyrae region. 

Analytical computations for the most relevant evolutionary
quantities have been published in the past by several authors 
\citep[see e.g.][]{Carretta2000, Chaboyer1998, Cassisi1998, buzzoni1983, 
Sweigart1978, Sweigart1976}.
However, these results were usually restricted to simple linear relations
among the evolutionary features of interest and some predictors
(or covariates), subsetting data at some fixed values of all other
predictors.   
Although a similar technique produces simple relations, these can not be
generalized to other values of the subsetting predictors.  
In the present work we chose an alternative multivariate approach,
allowing the regression models to include not only the predictors
but either their interactions. With this choice we were able to fit
the whole dataset with the same expression for all the values of predictors.  
We started with simple
relations including linearly the predictors but allowing for interaction between
chemical inputs ($Z$ and $Y$) and age (for TO luminosity) or mass of the
star. Then we checked whether the model was able to describe all significant
trends in the data without over-fitting them
\citep[see e.g.][]{linmodR}.
The first requirement was tackled by the analysis of the standardized
residuals, to check that the whole information present in the data was
extracted by  
the model. In this case the plot of standardized residuals versus the values
of the evolutionary feature predicted by the model should show the
points scattered without a clear path. 
Moreover, the plots of the standardized residuals versus the predictors were
used to infer the need to include quadratic or cubic therms or high-order
interactions. 
The plot of the standardized residuals also allows a visual check
of the hypothesis of homoscedasticity, i.e. that the
variance of the parent distribution of the residuals remain constant for
different values of the covariates -- or 
equivalently for different predicted values. 
The assessment of the statistical significance of the model covariates are
based upon the  
hypothesis that the parent distribution of the standardized residuals is the
standardized normal distribution $N(0,1)$.  
For a rough check of this hypothesis in the analysis we evaluated the quantile
2.5\%, 50\% and 97.5\% of the standardized residuals and compare them with
the corresponding quantile of $N(0,1)$. 
The problem of possible overfitting requires the use of the stepwise regression
\citep{venables2002modern} technique, which 
allows one to evaluate the performance of the multivariate model (balancing the
goodness-of-fit and the number of covariates in the model)
and of the
models nested in this one (i.e. models without some of the covariates).
To perform the stepwise model selection we employed the Akaike information
criterion (AIC): 
\[AIC = n \log \frac{d^2_E}{n} + 2 \; p \;, \]
which balances the number of covariates $p$ included in the model and its 
performance in the data description, measured by the error deviance $d^2_E$
($n$ is the number of points in the model). Among the models explored by 
the stepwise technique we selected that with the lower value of
the AIC as the best one.

To describe
the model concisely, in this section we used the
operator $*$, defined as 
$A * B \equiv A + B + A \cdot B$, and excluded the presence of the
regression coefficients in the models. An expanded version of all 
regression models is reported in Appendix \ref{app:models}.

For the turnoff luminosity we modeled the output of the
simulations with the following relation:
\begin{equation}
L_{\rm TO} = t_9 * (Y + \log Z) + \alpha_{ml} + {\cal K} \; ,
\label{eq:fit-to}
\end{equation}
where $t_9 = \log t$ ($t$ is the isochrone age in Gyr), $\alpha_{ml}$ is the
mixing length value. Since we explored the effect of only one possible 
$\alpha$-enhancement on the solar mixture, we chose to model its effect
by a categorical dicotomic variable $\cal K$. 
The model was fitted to the data with a least-squares method using the software R
2.13.1 \citep{R}.
The coefficients of the fit, along with their statistical significance, 
are listed in
Table~\ref{table:modTO}. In the first two columns of the table we report the
least-squares estimates of the regression coefficients and their errors; in the
third column we report the $t$-statistic for the tests of the statistical
significance of the covariates, and in the fourth column the
$p$-values of these tests.   

\begin{figure*}
\centering
\includegraphics[height=16.0cm,angle=270]{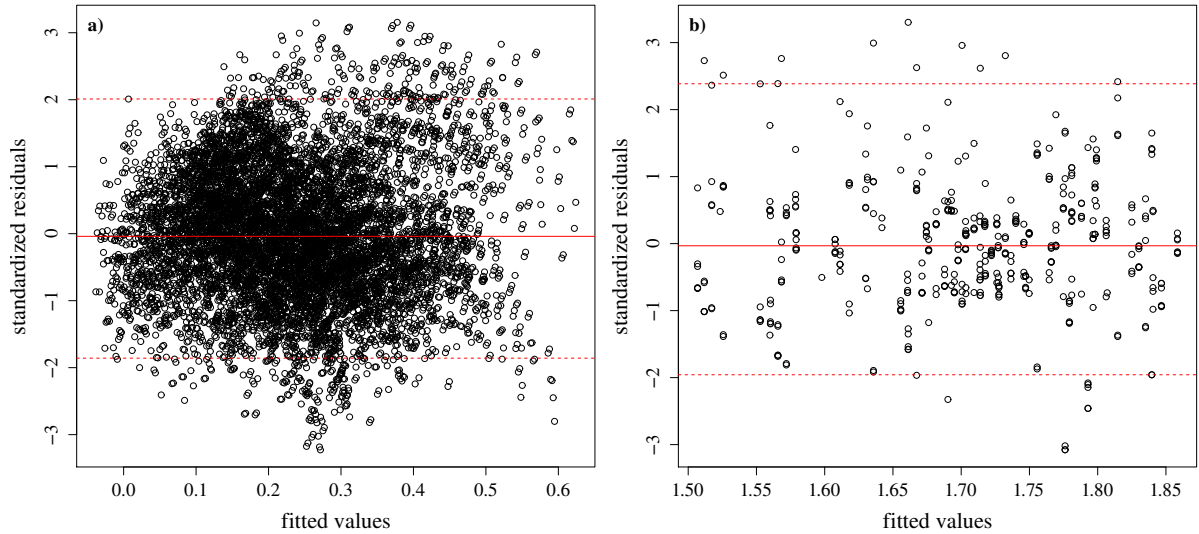}
\caption{Diagnostic plots showing standardized residuals of the fits versus
predicted values. Panel (a): $L_{\rm TO}$; panel (b): $L_{\rm HB-T3.83}$.
The dashed lines show the location of the quantiles
2.5\% and 97.5\% of the standardized residuals, while the solid one shows their
median value.}   
 \label{fig:anal-rel}
\end{figure*}

The residual standard error of the fit is $\sigma =
0.0088$, so that the fit is fairly accurate in the description of the data. 
The diagnostic plot of standardized residuals of the fit versus
predicted values is shown in Fig.~\ref{fig:anal-rel}, panel (a); it is
apparent that Eq.~\ref{eq:fit-to} gives a good analytical description of
the data. 
The effect of the $\alpha$-enhancement, although
statistically significant, is very small -- about 0.0019 dex -- and may be
safely neglected without modification in the model.

We calibrated of the ZAHB luminosity taken at $\log
T_{eff}=3.83$ in the central region of the RR Lyrae instability strip.
The luminosity was obtained by a linear interpolation in $\log T_{eff}$ 
on the ZAHB grid. 
We modeled the luminosity $L_{\rm HB-T3.83}$ with the following function
 of the mass of the star, the helium and the metal content:
\begin{equation}
L_{\rm HB-T3.83} = (M + M^2) * Y + \log Z \; ,
\end{equation}
where $M$ (in $M_{\sun}$) is the mass of the star.

However, the least-squares fit suffers of a heteroscedasticity problem 
because the 
plot of standardized residuals (not shown here) has a fan-shaped behavior,
showing an 
increase of the variance with the mass of the star.    
We modeled this
increase with a power law in the mass, $\sigma \propto M^{\beta}$, 
and corrected the eteroscedasticity by a weighted least-squares fit.
Through restricted maximum likelihood techniques we estimated the model
coefficients and the power $\beta$ that models the variance trend. We
performed the fit with the {\sl gls} function of the {\sl nlme} library
\citep{nlme} of the
R software. The model result is presented in Table \ref{table:modHB}.
The residual standard error of the fit is $\sigma = 0.032$, with
$\beta = 2.05$ (95\% confidence interval = [1.54, 2.55]). The diagnostic plot
in Fig.~\ref{fig:anal-rel}, panel (b) shows that the data are adequately
described by the model. 
\begin{table}
\centering
\caption{Fit of the turnoff luminosity.  In the first two columns: least-squares estimates of the regression coefficients and their errors;  
third column: $t$-statistic for the tests of the statistical 
significance of the covariates; fourth column: $p$-values of the tests. The
residual standard error is $\sigma = 0.0088$.} 
\label{table:modTO}      
\centering
\begin{tabular}{lrrrr}
\hline\hline
                   &   Estimate & Std. Error & $t$ value & $p$ value\\  
\hline  
(Intercept)        &  0.5964    & 0.008732   & 68.31   & $<2 \times 10^{-16}$\\
$t_9$              & -0.4972    & 0.008026   & -61.95  & $<2 \times 10^{-16}$\\
$Y$                & -1.150     & 0.01870    & -61.53  & $<2 \times 10^{-16}$\\
$\log Z$       & -0.3125    & 0.002006   & -155.80  & $<2 \times 10^{-16}$\\
$\alpha_{ml}$       &  0.03503   & 0.001165   & 30.07   & $<2 \times 10^{-16}$\\
$\cal K$ = AS09a3  &  0.001896  & 0.0001902  &  9.97   & $<2 \times 10^{-16}$\\
$t_9 \cdot Y$      & 0.3158     & 0.01770    & 17.84   & $<2 \times 10^{-16}$\\
$t_9 \cdot \log Z$  & 0.1731 & 0.001899  & 91.18   & $<2 \times 10^{-16}$\\
\hline
\end{tabular}
\end{table}

\begin{table}
\centering
\caption{Fit of HB luminosity taken at $\log
T_{eff}=3.83$. The residual standard error is $\sigma = 0.032$. The column
legend is the same as Table~\ref{table:modTO}.} 
\label{table:modHB}      
\centering
\begin{tabular}{lrrrr}
\hline\hline
 & Estimate & Std. Error & $t$ value & $p$ value\\
\hline
(Intercept)   & $-1.438$  & $0.4548$   & $-3.163$ & $1.7 \times 10^{-3}$\\
$M$           & $ 5.935$  & $1.136$    & $ 5.224$ & $2.7 \times 10^{-7}$\\
$M^2$         & $-3.481$  & $0.7425$   & $-4.688$ & $3.7 \times 10^{-6}$\\
$Y$           & $ 6.247$  & $1.283$    & $ 4.870$ & $1.6 \times 10^{-6}$\\
$\log Z$  & $-0.1067$ & $0.004814$ & $-22.17$ & $<2 \times 10^{-16}$\\
$M \cdot Y$   & $-11.14$  & $3.491$    & $-3.190$ & $1.5 \times 10^{-3}$\\
$M^2 \cdot Y$ & $ 5.844$ & $2.441$    & $ 2.394$ & $1.7 \times 10^{-2}$\\
\hline
\end{tabular}
\end{table}

\section{Conclusions}\label{sec:concl}

We presented a very large set of new stellar tracks and
isochrones computed with  
an updated version of the FRANEC code, which includes state-of-the-art input
physics  
(radiative and conductive opacity, equation of state, atmospheric models and
nuclear cross-sections).  
The main novelties of these models with respect to those currently available
in the literature are the adoption  
of the heavy-element solar mixture by \citet{AGSS09}, the recent
$^{14}N(p,\gamma)^{15}O$ reaction rate 
 by \citet{14n}, and the boundary conditions from detailed atmosphere models. 
 
With the aim to provide a powerful and versatile tool for the interpretation
of the unceasingly growing  
amount of data, we computed a very large database covering a fine grid of
masses, ages, and chemical compositions.  
More specifically, in the mass range $0.30 \div 1.10$ $M_{\sun}$, we made
evolutionary tracks and isochrones for 19 
metallicities available, ranging from $Z$ = 0.0001 to 0.01, 
and five different helium abundances for each $Z$ ranging from $Y$ = 0.25 to
0.42. The availability  
of sets of models with initial helium abundance as high as 0.33, 0.38,
and 0.42 is of  
 primary importance in the context of multipopulation globular cluster studies. 
For each choice of initial metallicity $Z$ and helium abundance $Y$, we
computed tracks and isochrones  
with two different element mixtures, namely solar-scaled by \citet{AGSS09} and
$\alpha$-enhanced with [$\alpha$/Fe] = 0.3. 
Finally, we provided all these sets of models for three different values of
the mixing-length parameter $\alpha_{ml}$ = 1.70, 1.80, and 1.90. 
Each set contains evolutionary tracks from the pre-MS to the helium flash, HB
models, 
and isochrones in the age range $8 \div 15$ Gyr, in time steps of 0.5 Gyr.
 
The database, currently consisting of about 33000 stellar tracks and about 10000
isochrones, is available on the
web\footnote{\url{http://astro.df.unipi.it/stellar-models/}}.  

Models were compared with other computations available in the literature and
with data of selected globular clusters. 

We also provided useful analytical relations describing the dependence of
relevant evolutionary quantities, namely turn-off and
horizontal branch  
luminosities, on the chemical composition and convection
efficiency.  
More important, we analyzed these relations for the first time in a thorough statistical way to obtain simple but accurate models and to
shed some light on the interesting interactions of the chemical and physical
inputs of the simulations.

\begin{acknowledgements}
We are grateful to G. Bono, A. Di Cecco, and P. Stetson, who kindly provided us
with the globular cluster observational data. 
We wish to thank our anonymous referee for very useful comments and
suggestions.  
It is a pleasure to thank Emanuele Tognelli, Rosa
Becucci, Federica Zacchei for useful and pleasant discussions. We thank Steve
Shore for a careful reading of the manuscript. 
This work has been supported by PRIN-INAF 2008 (P.I. Marcella Marconi).
\end{acknowledgements}
\bibliographystyle{aa}
\bibliography{biblio}

\appendix
\section{Full form of the regression models}\label{app:models}

The full form of the proposed model for the TO luminosity is
\begin{eqnarray}
L_{\rm TO} & = & \beta_0 + \beta_1 \; t_9 + \beta_2 \;Y + \beta_3 \; \log Z  +
\beta_4 \; \alpha_{ml} + \beta_5 \; {\cal K} + \nonumber \\
         & + & \beta_6 \; t_9 \cdot Y + \beta_7 \; t_9 \cdot \log Z \;,
\end{eqnarray}
the regression coefficients $\beta_i$ are listed in the same order as in
Table~\ref{table:modTO}.

The full form of the model of the luminosity $L_{\rm HB-T3.83}$ is
\begin{eqnarray}
L_{\rm HB-T3.83} & = & \beta_0 + \beta_1 \; M + \beta_2 \; M^2 + \beta_3 \; Y
+ \beta_4 \; \log Z + \nonumber \\
 & + & \beta_5 \; M \cdot Y +  \beta_6 \; M^2 \cdot Y \;,
\end{eqnarray}
the regression coefficients $\beta_i$ are listed in the same order as in
Table~\ref{table:modHB}.

\end{document}